\documentstyle[12pt,aps,prl]{revtex}
\tighten
\begin{document}
\author{G. Dautcourt}  
\address{Max-Planck Institut f\"ur Gravitationsphysik, 
Albert-Einstein-Institut \\
Schlaatzweg 1, D~--~14473 Potsdam, Germany 
}
\title{On the Ultrarelativistic Limit of General Relativity}
\maketitle
\begin{abstract}

As is well-known, Newton's gravitational theory can be formulated as a 
four-dimensional space-time theory and follows as singular limit from 
Einstein's theory, if the velocity of light tends to the infinity. Here 
'singular' stands for the fact, that the limiting geometrical 
structure differs from a regular Riemannian space-time. Geometrically, 
the transition Einstein $\rightarrow$ Newton can be viewed as an 
'opening' of the light cones. This picture suggests that there might 
be other singular limits of Einstein's theory: Let all light cones 
shrink and ultimately become part of a congruence of singular world 
lines. The limiting structure may be considered as a nullhypersurface 
embedded in a five-dimensional spacetime. While the velocity of light 
tends to zero here, all other velocities tend to the velocity of 
light. Thus one may speak of an ultrarelativistic limit of General 
Relativity. The resulting theory is as simple as Newton's 
gravitational theory, with the basic difference, that Newton's 
elliptic differential equation is replaced by essentially ordinary 
differential equations, with derivatives tangent to the generators of 
the singular congruence. The Galilei group is replaced by the Carroll 
group introduced by L\'evy-Leblond. We suggest to study near 
ultrarelativistic situations with a perturbational approach 
starting from the singular structure, similar to post-Newtonian 
expansions in the $c \rightarrow \infty$ case.
\end{abstract}
\section{Introduction}
\label{intro}
General Relativity (GR) not only governs the gravitational interactions
between bodies, it also dictates the causality structure of spacetime.
This latter property is most interesting, if the gravitational
fields become strong and develop singularities. 
In a limit, when the whole spacetime becomes singular or nearly singular, 
the causality structure should strongly deviate from that of a
near-Minkowskian geometry. 
It is one of the virtues of GR, that the theory covers 
- if properly interpretet - even such extreme situations.
 
A well-known example is Newton's theory of gravity. 
Its four-dimensional formulation requires a spacetime structure, 
which is singular from the viewpoint of Riemannian geometry 
\cite{Cartan}. Using Einstein's field equations, this singular 
structure is obtained, if the velocity of light is taken to tend to 
infinity \cite{Newton}. Geometrically, the transition 
Einstein $\rightarrow$ Newton can be viewed as an opening of the 
light cones. In the limit $c \rightarrow \infty$, the cones 
become the spacelike hypersurfaces of Newton's absolute time. 
In spite of Kuhns claims about the incommensurability of 
concepts of successive theories, both Newton's and Einstein's 
theory can be covered by a common spacetime theory \cite{Ehlers}.
Nevertheless, the causality structure of Newton's theory is radically
different from that of Einstein's: Interactions occur simultanously on
the hypersurfaces of constant Newtonian time ("action at a distance").
This is reflected by the existence of an {\it elliptic} differential 
equation for the Newtonian potential, as compared to {\it hyperbolic}
differential equations for time-dependent situations in GR. 
Closely related is the fact that the Poincar\'e group is replaced 
by the Galilei group.

The visualization of the transition Einstein $\rightarrow$ Newton 
suggests immediately, that Newton's theory may not be the only 
singular limit of Einstein's theory. We here discuss a situation, 
which is in some sense opposite to the Newtonian case. Let all 
light cones shrink and ultimately become part of a congruence of 
singular world lines. Geometrically, this limiting structure may be 
considered as a four-dimensional nullhypersurface $V^{(1)}_{4}$ 
embedded in a five-dimensional spacetime. 
While the light velocity tends to zero here, all other 
velocities tend to the velocity of light. One may therefore speak of 
an ultrarelativistic limit of GR (see \cite{Daut69},\cite{Henneaux},
\cite{Levy} for previous discussions). 
Again, the causality structure in the limit is different: 
Instead of the hyperbolic differential equations of GR and elliptic 
differential equations of Newton's theory, we have now essentially 
{\it ordinary} differential equations, with derivatives tangent to 
the generators of the singular congruence. There are no interactions 
between spatially separated events, and no true motion occurs in the 
limit, except for tachyonic motion. However, the isolated and immobile 
physical objects show evolution. A situation of this
type has sometimes been called "Carroll causality" \cite{Henneaux}, 
after Louis Carroll's tale Alice in Wonderland. It may also be 
characterized as ultralocal approximation, which is perhaps a better
notation than ultrarelativistic. 
While the Newtonian limit is governed by 
the Galilei group, the
invariance group in the ultrarelativistic limit is another degenerate 
limit of the Poincar\'e group, the Carroll group, introduced by 
L\'evy-Leblonc, who also first discussed its representations and 
its Lie algebra \cite{Levy}.

One expects, that an ultrarelativistic approximation procedure, 
starting from the singular spacetime and similar 
to the method considered in the opposite Newtonian case \cite{Daut97}, 
might be useful for situations of strong gravity. 
It is encouraging that the field equations in the ultrarelativistic 
limit are as simple as in Newton's theory. 

In section 2 the geometry of 
a stand-alone singular spacetime $V_4^{(1)}$ is considered, 
independently of the limiting procedure and of 
its embedding into higher-dimensional spaces. A Riemannian 
curvature tensor based on second-order derivatives of the 
metric is not a genuine geometrical construction here, since
no uniquely defined intrinsic connection exists. However,   
Ricci rotation coefficients can be introduced.  
The use of adapted coordinates simplifies the 
relations. Section 3 considers a family 
$g_{\mu\nu}(x^{\mu},\epsilon)$ of metrics, 
satisfying the general-relativistic field equations and tending 
for $\epsilon ~(= c^2) \rightarrow 0$ to the singular spacetime 
introduced in the previous section. The resulting ultrarelativistic 
field structures depend on the type and behaviour of matter fields 
for $\epsilon  \rightarrow  0$, which are present together with the
gravitational field. In section 4 some solutions of the 
ultrarelativistic field equations are discussed,
including pure vacuum gravitational fields and dust matter.

Many problems remain open. The limit discussed here can be 
considered for any general-relativistic 
field theory. It is straightforward to set up a 
post-ultrarelativistic expansion and thus to re-introduce 
the velocities which have disappeared in the limit.  
Another open question is the relation of solutions of the 
ultrarelativistic equations to those general-relativistic
solutions, which admit an ultrarelativistic limit. 

\section{Differential geometry of singular Riemannian spaces 
$V^{(1)}_{4}$} 
\label{V5}

A degenerate Riemannian space $V^{(m)}_n$ may be defined as a
$n$-dimensional Riemannian space equipped with a covariant metric 
tensor $\gamma_{\mu\nu}$ of matrix rank $0<m<n$ ~\cite{Daut68}. 
Well-known examples are the usual nullhypersurfaces with $n=3$ and 
$m=1$ (see, e.g., \cite{Daut67}). We are interested in the case 
$n=4, m=1$.  At any given point the metric tensor can be reduced to 
$\gamma_{0\mu}=0, \gamma_{ik}=\delta_{ik}$ by means of suitable 
coordinate transformations. The values are preserved under the 
transformation $x'^0= g(x^0), x'^i= R^i_kx^k+ S^i$, where $R^i_k$ is
a 3-rotation and $g$ an arbitrary function of $x^0$. The linear subgroup 
of this transformation group is the 10-parameter Carroll group 
\cite{Levy}. Returning to a general coordinate system, at every point 
there exists a contravariant vector field $k^\mu 
~(\mu = 0,...3)$ which is a nonvanishing solution of 
\begin{equation}
\gamma_{\mu \nu}k^\nu = 0,
\end{equation}
defined up to an arbitrary factor. Furthermore, there exists  
a congruence of curves $x^\mu= x^\mu(\xi^i,v),~i=1,2,3$, called 
generators of $V_4^{(1)}$, to which the directions $k^\mu$ are 
tangent, as solutions to the differential equation 
\begin{equation}
k^\mu(x^\mu) = \frac{\partial x^\mu}{\partial v}.
\end{equation}
The three quantities $\xi^i$ fix a generator, and the parameter $v$
along a generator is determined up to a transformation 
$v' = v'(v,\xi^i), \frac{\partial v'}{\partial v} \neq 0$. $k^\mu$
is complemented by three other contravariant vectors $l^\mu_{(i)}$
such that $\gamma_{\mu\nu}l^\nu_i \neq 0$ and 
\begin{equation}
\gamma_{\mu\nu}l^\mu_{(i)}l^\nu_{(k)} = \delta_{ik}.
\end{equation}
The four vectors $(k^\mu, l^\mu_{(i)})$ form a contravariant tetrad, 
spanning the tangent
space at every point of the $V_4^{(1)}$. The cotangent space is 
spanned by the three vectors 
\begin{equation}
l_{\mu(i)}= \gamma_{\mu \nu}l^\nu_{(i)} \neq 0,
\end{equation}
and  
\begin{equation}
k_\mu = 
\frac{\epsilon_{\mu\nu\rho\sigma} l^\nu_{(1)}l^\rho_{(2)}l^\sigma_{(3)}}
{\epsilon_{\alpha\beta\gamma\delta}
k^\alpha l^\beta_{(1)}l^\gamma_{(2)}l^\delta_{(3)}},
\end{equation}
where $\epsilon_{\mu\nu\rho\sigma}$ is the Levi-Civita density. Note 
\begin{equation}
l^\mu_{(i)}l_{\mu (k)} =  \delta_{ik}, ~k^\mu l_{\mu(i)} =0, 
\end{equation}
\begin{equation}
l^\mu_{(i)}k_\mu  = 0, ~k^\mu k_\mu = 1. 
\end{equation}
The metric is written as $\gamma_{\mu\nu}= l_{\mu(i)}l_{\nu(i)}$,
and the tetrad is determined up to the generalized four-dimensional 
null rotations
\begin{eqnarray}
l'^{\mu}_{(i)} & =& A^k_{i}l^\mu_{(k)} + B_ik^\mu,  \\ 
k'^{\mu} &=& C k^\mu,   
\end{eqnarray}
which form a 7-parameter group, the coefficients $A^k_i$
represent a 3-rotation. There exists no contravariant metric 
tensor $\epsilon^{\mu\nu}$ satisfying
$\epsilon^{\mu\rho}\gamma_{\nu\rho}= \delta^\mu_\nu$, however 
$\gamma_{\mu\rho}\gamma_{\nu\sigma}\epsilon^{\rho\sigma}
= \gamma_{\mu\nu}$ has solutions. The simplest one is given by 
\begin{equation}
\epsilon^{\mu\nu}= l^\mu_{(i)}l^\nu_{(i)}, 
\end{equation}
but depends on the choice of the tetrad. One can also easily show,
that in general there exists no connection $\Gamma^\rho_{\mu\nu}$,
satisfying the Ricci lemma $\gamma_{\mu\nu;\rho}= 0$ and depending 
only on the metric and its {\it first} derivatives. Instead, one 
may define tetrad-dependent affine connections by
\begin{equation}
\Gamma^\rho_{\mu\nu} = 
\frac{1}{2} k^\rho(k_{\mu,\nu} + k_{\nu,\mu})
+ \epsilon^{\rho\sigma}\Gamma_{\mu\nu\sigma} \label{aff}
\end{equation}
with 
\begin{equation}
\Gamma_{\mu\nu\rho}= 
\frac{1}{2}(\gamma_{\rho\mu,\nu}
+\gamma_{\rho\nu,\mu}-\gamma_{\mu\nu,\rho}).
\end{equation}
The affine connection (\ref{aff}) also does not in general satisfy 
the Ricci lemma $\gamma_{\mu\nu;\rho}=0$, one obtains instead
\begin{equation}
\gamma_{\mu\nu;\rho}= h_{\mu\rho}k_\nu + h_{\nu\rho}k_\mu, 
\end{equation}
where the tensor 
\begin{equation}
h_{\mu\nu}= \Gamma_{\mu\nu\rho}k^\rho 
\end{equation}
is (up to a factor $\frac{1}{2}$) the Lie derivative of the 
metric in the direction $k^\rho$.
The $\Gamma^\rho_{\mu\nu}$ transform as an affine connection with 
respect to coordinate transformations, thus the four-dimensional 
Ricci and Riemann tensors formed with the connection are indeed 
tensors. They have nevertheless no geometrical meaning in general, 
since they depend on the choice of tetrad, and are therefore to an 
large extent arbitrary. The situation can be different for 
nullhypersurfaces $V_3^{(1)}$ embedded in a Riemannian $V_4$. 
Here a rigging of the surfaces would allow us to fix the affine 
connection and to introduce tensorial curvature measures (for 
different geometries on nullhypersurfaces, see \cite{Daut67} and 
\cite{Penrose}). A way to obtain true geometrical statements in 
$V_4^{(1)}$ is the introduction of Ricci rotation coefficients, 
which are obtained by expressing the derivatives of the tetrad 
$(k^\mu, l^\mu_{(i)})$ in terms of tetrad \cite{Daut68}. 
The rotation coefficients are scalars with respect to coordinate 
transformations, but transform under a tetrad change. 
Geometrically relevant propositions may be formulated as 
tetrad-invariant statements on the rotation coefficients. For
instance, differential invariants of the $V_4^{(1)}$ may
be given as suitable functions of the rotation coefficients 
and their derivatives. Contrary to nonsingular Riemannian 
geometries, invariants depending only on the metric and its {\it first}
derivatives exist here, {\it six} for the 
$V_4^{(1)}$, {\it one} for the $V_3^{(1)}$.
It is possible to write down the invariants and field equations in 
the $V_4^{(1)}$ in a {\it manifestly covariant} manner. However, 
as in Newtonian theory, the existence of intrinsic geometrical 
structures allows the introduction of adapted coordinates, here $\xi^i$ 
and $v$. Since adapted coordinates simplify many relations 
considerably, they will be used throughout subsequently. Note 
that they are determined up to the transformations $v'=v'(v,\xi^i),
 ~\xi'^{i}=\xi'^{i}(\xi^i)$.

\section{The transition $c \rightarrow 0 $}
\label{transition}
We assume that the singular space $V_4^{(1)}$ arises as the limit 
$\epsilon \rightarrow 0$ of a family of normal-hyperbolic Riemannian 
spacetimes, with the metric $g_{\mu\nu}(x^{\mu},\epsilon)$,
satisfying the Einstein field equations for $\epsilon > 0$. $\epsilon$
is taken as the {\it square} of the velocity of light, $\epsilon=c^2$.
A mathematically rigourous approach should employ Geroch's technique of 
embedding in a 5-manifold \cite{Geroch}. Instead, we use for simplicity 
asympotic representations for the metric, writing down an expansion 
of the type
\begin{equation} 
g_{\mu\nu}(x^\mu,\epsilon) = \gamma_{\mu\nu}(x^\mu) 
+ g_{(1)\mu\nu}(x^\mu)\epsilon +g_{(2)\mu\nu}(x^\mu)\epsilon^2 
+o(\epsilon^3)
\label{gd}
\end{equation}
Assuming $g_{(1)\mu\nu}k^\mu k^\nu \neq 0$, the contravariant 
components of the metric may be represented asymptotically as 
\begin{equation}
g^{\mu\nu}(x^\mu,\epsilon) = \frac{1}{\epsilon}fk^\mu k^\nu 
+  g_{(0)}^{\mu\nu}(x^\mu)+ g_{(1)}^{\mu\nu}(x^\mu)\epsilon + o(\epsilon^2),
\label{gu}
\end{equation}
where $f(x^\mu)$ is a scalar function. In adapted coordinates 
we put $k^\mu = \delta^\mu_0$. The relations between the co- and
contravariant metric components give
\begin{equation}
g_{(1)00}= 1/f, ~g_{(1)0k}= -\gamma_{ik}g_{(0)}^{0i}/f, 
~g_{(0)}^{ik}= \gamma^{ik} 
\end{equation}
($\gamma^{ik}$ is inverse to $\gamma_{ik}$).
It is useful to expand also the coordinate transformation as a 
series in $\epsilon$:
\begin{equation}
x'^\mu(x^\rho) = x_{(0)}^\mu(x^{\rho})
+ \epsilon x_{(1)}^{\mu}(x^\rho) + o(\epsilon^2).
\end{equation}
Putting $x'^{\mu}_{(0)} =x^\mu$, the next order $x_{(1)}^\mu$ 
can be considered as gauge. From 
\begin{eqnarray}
g'^{0i}_{(0)} & =& g_{(0)}^{0i} +f\frac{\partial x'^i_1}{\partial v},\\
g'^{00}_{(0)} & =& g_{(0)}^{00} +2f\frac{\partial x'^0_1}{\partial v},\\
g'^{ik}_{(0)} & =& \gamma^{ik}
\end{eqnarray}
it is evident, that $g_{(0)}^{0i}$ and $g_{(0)}^{00}$ may be 
transformed to zero. This simplifies the field equations considerably. 
If we calculate the Ricci tensor with the series (\ref{gd}),(\ref{gu}), 
singular terms proportional to $\epsilon^{-2}$ and $\epsilon^{-1}$ 
arise. A closer inspection shows that $R_{(-)2\mu\nu} = 
\lim_{\epsilon \rightarrow 0} (\epsilon^2 R_{\mu\nu})$ 
reduces to zero. We use the field equation with a matter tensor 
$T_{\mu\nu}$. The product of $T_{\mu\nu}$ with $\kappa = 
 \frac{8\pi G}{c^2}$ is assumed to have for $\epsilon \rightarrow 0$ 
a finite limit $t_{0\mu\nu}$, which allows us to write
\begin{equation}
\kappa T_{\mu\nu}(x^\mu,\epsilon) = 
t_{(0)\mu\nu}+ t_{(1)\mu\nu}\epsilon
  + o(\epsilon^2). 
\end{equation}
It is not difficult to show that these assumptions are compatible with,
e.g., dust matter.
Then the field equations start with 
\begin{eqnarray}
R_{(-1)\mu\nu} 
=- \frac{1}{2}g_{(0)\mu\nu}t_{(0)00} f, 
  \label{minus}\\
R_{(0)\mu\nu}  = t_{(0)\mu\nu}  -\frac{1}{2}g_{(1)\mu\nu}t_{(0)00}f 
\nonumber \\
-\frac{1}{2}g_{(0)\mu\nu}( t_{(0)\alpha\beta}g^{(0)\alpha\beta}
+t_{(1)00}f).                  
\label{zero}
\end{eqnarray}
From equation (\ref{minus}) only the pure spatial components survive. 
The ultrarelativistic field equations are obtained from (\ref{minus}) 
(see below (\ref{prop}) and (\ref{dprop})) and from the time-time and 
time-space components of (\ref{zero}) (below (\ref{in1}),(\ref{in2}) 
and (\ref{din1})). The space-space components of (\ref{zero}) introduce 
already post-ultrarelativistic corrections, which are not discussed 
here. 

\section{Some solutions}
\label{sol}
The vacuum field equations can be written with $f=-e^{-H}$
\begin{equation}
\dot{\gamma}_{ik}\dot{\gamma}^{ik} +(\dot{\gamma}_{ik}\gamma^{ik})^2 =0, 
\label{in1}
\end{equation}
\begin{equation}
\gamma^{kl}\dot{\gamma}_{kl|i}- \gamma^{kl}\dot{\gamma}_{il|k} 
- \frac{1}{2}H_{,i}\gamma^{kl}\dot{\gamma}_{kl} = 0 \label{in2}
\end{equation}
and
\begin{equation}
\ddot{\gamma}_{ik} 
+ \frac{1}{2}\dot{\gamma}_{ik}(\gamma^{lm}\dot{\gamma}_{lm}
-\dot{H})- \dot{\gamma}_{il}\dot{\gamma}_{km}\gamma^{lm} =0, 
\label{prop}
\end{equation}
where a stroke denotes the covariant derivative with respect to the 
3-metric $\gamma_{ik}$, a dot means $\partial/\partial v$ ('time'-derivative). 
These equations should describe the motion 
of gravitational waves in the ultrarelativistic limit. One recognizes 
an initial value problem with (\ref{in1}),(\ref{in2}) as initial 
conditions and (\ref{prop}) as propagation equation. Equation 
(\ref{in1}) (which signifies the vanishing of one of the first-order 
invariants of the $V_4^{(1)}$ mentioned above) is preserved under 
(\ref{prop}), but (\ref{in2}) leads to the additional constraint
\begin{equation}
\dot{\gamma}_{ik}\gamma^{kl}\dot{H}_{,l} = 0. \label{in3}
\end{equation}
Hence, considering only the general case 
$det ~|\dot{\gamma}_{ik}\gamma^{kl}| \neq 0$, $H$ is given by 
\begin{equation}
 H = h(v) + hh(\xi^i),
\end{equation}
where $hh(\xi^i)$  enters the initial conditions (\ref{in2}) and $h(v)$ 
influences the propagation equation (\ref{prop}). As an example
consider the vacuum solution
\begin{equation}
\gamma_{ik} = diag(\gamma_1,\gamma_2,\gamma_2),~H= -4 ln(a+bv) 
+hh(\xi^1)
\end{equation}
with 
\begin{equation}
\gamma_1 = a+bv, \gamma_2\gamma_3 = c(\xi^2,\xi^3)e^{-hh(\xi^1)}.
\end{equation}
Time-dependent solutions of this type may be considered as the 
ultrarelativistic limit of gravitational waves. It is seen that 
caustics of the congruence may occur at $v=-a/b$ in the example.

The case of dust matter is only slightly more complicated. 
The propagation equation for the 3-metric attains a source term 
(writing $\rho$ for the $c \rightarrow 0$ limit of the matter density) 
\begin{equation}
\ddot{\gamma}_{ik} 
+ \frac{1}{2}\dot{\gamma}_{ik}(\gamma^{lm}\dot{\gamma}_{lm}
-\dot{H})- \dot{\gamma}_{il}\dot{\gamma}_{km}\gamma^{lm} =
8\pi G \rho e^{H}\gamma_{ik}. \label{dprop}
\end{equation}
A source term is also present in the first (scalar) initial equation
\begin{equation}
\dot{\gamma}_{ik}\dot{\gamma}^{ik} +(\dot{\gamma}_{ik}\gamma^{ik})^2 
= 64\pi G \rho e^{H}. \label{din1}
\end{equation}
The second (vectorial) initial equation (\ref{in2}) remains unchanged.
The scalar constraint (\ref{din1}) is preserved in time, 
if the matter density $\rho$ changes as 
\begin{equation}
\frac{\dot{\rho}}{\rho} = -\frac{1}{2}\dot{\gamma}_{ik}\gamma^{ik}.
\end{equation}
The latter relation is equivalent to 
$\rho\sim (det |\gamma_{ik}|)^{-1/2}$,
which corresponds to matter conservation in comoving coordinates 
(note that $\rho$ can vary arbitrarily as a function of $\xi^i$).  
The vectorial constraint equation (\ref{in2}) is not preserved 
in time, but leads to the relation
\begin{equation}
\dot{\gamma}_{ik}\gamma^{kl}\dot{H}_{,l} 
+ \frac{3}{8}H_{,i}(\dot{\gamma}_{kl}\dot{\gamma}^{kl} 
+ (\dot{\gamma}_{kl}\gamma^{kl})^{2}) =0.
\end{equation}
Again, the integration of these equations is much simpler than the 
corresponding general relativistic equations. Consider the general 
case of an isotropic expansion, defined by 
\begin{equation}
\dot{\gamma}_{ik} = \phi(\xi^i,v)\gamma_{ik}.
\end{equation}
The solution can be written in terms of an arbitrary function 
$\lambda(v)$ as
\begin{equation}
\phi(\xi^i,v)= 
\frac{\lambda}{r_0^3(1+\frac{9}{4r_0^3}\int^v_{v_0}\lambda \,dv)}.
\end{equation} 
The arbitrariness of $\lambda(v)$ reflects the fact, that no affine
parameter along the generators is singled out, all parameters 
$v'=v'(v)$ are at the same footing. $r_0$ is a spatially varying 
length scale: Assuming an initial density $\rho_0(\xi^i)$ at $v=v_0$, 
we define this scale by
\begin{equation}
r_0(\xi^i)= (\frac{3}{32\pi G\rho_0(\xi^i)})^{1/2}. 
\end{equation}
The matter density as a function of $v$ is then given by
\begin{equation}
\rho(\xi^i,v) 
= \frac{\rho_0(\xi^i)}{(1+\frac{9}{4r_0^3}\int^v_{v_0}\lambda\,dv)^{2/3}},
\end{equation}
and $H$ can be found from $e^H =3\phi^2/(32\pi G\rho)$.
It is interesting, that a singular origin of the expanding
matter distribution is as inevitable as in GR. 

\section*{Acknowledgments}
I appreciate useful remarks from and discussions with many participants 
of the Jadwisin workshop. In particular I wish to thank 
P. Aichelburg, C. B\"ar, S. Bazanski, W. Drechsler, J. Ehlers, 
J. Goldberg, F. Hehl, H. Kleinert, C. L\"ammerzahl, Y. Ne'eman, 
J. Stachel, A. Trautman and H. Urbantke. 

\end{document}